\documentclass[12pt]{article}
\usepackage{amssymb}
\usepackage{amsfonts}

\usepackage[all]{xy}

\begin{document}

\title{Automaton model of protein: dynamics of conformational and functional states}

\author{Andrei Khrennikov and Ekaterina Yurova\\ International Center for Mathematical Modeling\\
in Physics and Cognitive Sciences\\
Linnaeus University, V\"axj\"o, S-35195, Sweden\\
Andrei.Khrennikov@lnu.se, Ekaterina.Yurova@lnu.se}

\maketitle

\begin{abstract}
In this conceptual paper we propose to explore the analogy between ontic/epistemic description of quantum phenomena and 
interrelation between dynamics of conformational and functional states of proteins. Another new idea is to apply theory of  
automata to model the latter dynamics. In our model protein's behavior is modeled with the aid of two dynamical systems, 
ontic and epistemic, which describe evolution of conformational and  functional states of proteins, respectively. The epistemic 
automaton is constructed from the ontic automaton on the basis of functional (observational) equivalence relation on the space of ontic states.
This reminds  a few approaches to emergent quantum mechanics in which a quantum (epistemic) state is treated as representing 
a class of prequantum (ontic) states.  This approach does not match to the standard {\it protein structure-function paradigm.} 
However, it is perfect for modeling of  behavior of intrinsically disordered proteins. Mathematically space of protein's ontic states
(conformational states) is modeled with the aid of $p$-adic numbers or more general ultrametric spaces encoding the internal 
hierarchical structure of proteins. Connection with theory of $p$-adic dynamical systems is briefly discussed.    
\end{abstract}

keywords: proteins,  conformational and functional states, structure-function paradigm, automaton-model, quantum-like model

\section{Introduction}

In this note we apply automata theory \cite{Sakarovitch}, \cite{A4a}, \cite{A4}   to modeling dynamics of proteins' states. The protein automaton model is presented by exploring  the  scientific methodology which 
plays the important role  in foundations of quantum physics. This methodology is based on joint exploring of the two descriptive levels: {\it ontic and epistemic.} The ontic description 
is about reality ``as it is'' and the epistemic description is about observations, see Atmanspacher et al.  \cite{H1}-\cite{H4}, see also \cite{KHR_PRS1}-\cite{KHR_PRS3}.
Now we want to apply this methodology to molecular biology. Conformation 
states of proteins are treated as ontic states and functional states or states approachable for observations as epistemic states. Such a separation of two descriptive levels is very 
fruitful in creation of an adequate model of dynamics of proteins' states. Moreover, it matches well with automata theory \cite{Sakarovitch}: automata without output represent dynamics of 
ontic states, in our case conformational states, and automata with outputs represent dynamics of epistemic states, in our case functional/observational states.    

 We do not pretend that proteins are genuine  quantum systems; we just want to use the methodology of ontic-epistemic states designed for quantum mechanics, see appendix 1 for discussion. 
The starting point linking dynamics of proteins  with factorization models of quantum phenomena is the observation that the space of conformational states 
of a protein is huge and the concrete conformational states can be 
unobservables. Thus, as in quantum physics, one  has to proceed with two types of states: ontic states (conformational states of a protein) 
and epistemic states (observable states representing various 
functions of a protein).  We remark that, for a protein, the number of epistemic states is small, since each protein is a highly specialized biological machine with a few functions.
At the same time the number of conformational states is huge. 

In short we plan to model protein's state dynamics as functioning of an automaton without output having a huge configuration space (of conformational states); by adding output to this automaton 
we construct the model of observations of protein's functional states (epistemic states). Such an observational output produces factorization of the space of conformational states and 
can be mathematically represented as new ``factorized automaton''. In the framework of this automaton-model we discuss a number of conceptual problems related to production 
of functional states of proteins from conformational states.  {\it We emphasize that this is a concept-type paper.} Our main aim is to present the concept of {\it the automaton-protein.} 
We also rise a series of questions which have to be clarified to make this model closer to the real biological situation, section \ref{PRB}. We plan to work on these questions in coming publications.

 This model of protein's behavior does not match  the standard {\it protein structure-function paradigm} formulated  more than one hundred years ago \cite{Fischer}, \cite{SFP}. 
(In the operational terms by this paradigm the ontic and epistemic descriptions can be unified in the straightforward manner.) At the same time the
ontic/epistemic operational approach  is perfect for modeling of  behavior of intrinsically disordered proteins \cite{IDP1}, see section \ref{MB} for the discussion on recent studies
of the molecular biological features of proteins' functioning.    

We want that the paper will be readable for biologists. Therefore we try to minimize the mathematical apparatus in the use. The basic mathematical formalism under 
consideration is theory of {\it automata}, see \cite{Sakarovitch} and section \ref{AT} for definition.  We illustrate our model by a few illustrative examples, see Figures 1-3. 

The presented model can be formulated in more advanced mathematical terms; in particular, by formalizing the mathematical structure of the ontic state space. We discuss 
shortly this problem appendix 2.  The most natural model of proteins' state space is based on so called $p$-adic numbers, where $p > 1$ is a natural number. For $p$ prime, these numbers are 
widely used in theory of dynamical systems,  physics, biology and psychology \cite{A1}-\cite{GI2}.  We remark  a kind of similarity between the $p$-adic state 
representation of protein's state dynamics and application of $p$-adic dynamical systems to encryption of data, see Anashin \cite{A1, A2, A3}. Thus one may speculate 
that nature encrypted functioning of proteins by using methods of $p$-adic cryptography. This is an interesting idea, but in this paper we cannot elaborate it in more detail 
(in particular, such elaboration would lead to exploration of advanced mathematics of theory of $p$-adic dynamical systems).   

Finally, we remark that statistics of quantum observations has some peculiarities  \cite{KHR_WSP}, see also appendix 1 with discussion on the Bell inequality.
 In physics it is known that automata-dynamics can generate  some features of quantum statistics \cite{H1}. In  this note we show that 
biological systems such as proteins (and others having huge configuration space of internal 
states and just a few functional/observational  states) can demonstrate special statistical features (even generation of $p$-adic probabilities corresponding to the absence 
of statistical stabilization of frequencies of outputs, see \cite{KHR_P1},  \cite{KHR_P2}  for such probabilities and their role in non-Archimedean physics and biology).        

\section{Dynamics of proteins conformational and functional states from the perspective of molecular biology}
\label{MB}

In this section we apply the ontic-epistemic approach  to state dynamics (which we borrow from quantum foundations) to the protein dynamics. 

By the   {\it the protein structure-function paradigm,} any protein function is determined by a fixed three-dimensional structure \cite{SFP}.  This paradigm has been dominating 
in molecular biology during more than one hundred years, see Fischer \cite{Fischer} for one of the pioneer studies. At the same time it was challenged 
during many years including experimental evidences of its violation. Nevertheless, it stayed firmly as the cornerstone of molecular biology; only the discovery of 
{\it intrinsically disordered proteins} (IDPs),  Dunker et al. \cite{IDP1},   Dyson and Wright \cite{IDP2}, Dunker et al. \cite{IDP3}   changed (at least partially) the common viewpoint on the structure-function isomorphism.  
Nowadays it is widely understood that {\small ``functioning proteins do not remain fixed in a unique structure, but instead they sample a range of conformations facilitated by
motions within the protein'',} Ramanathan et al. \cite{PR}, see also Henzler-Wildman et al. \cite{PT}. 

We recall that the {\it native state} of a protein is its completely folded functional state.   However, opposite to the protein structure-function paradigm, 
the native state is no uniquely determined by protein's geometry. We again cite Ramanathan et al.  \cite{PR}:

{\small ``Even in the native state, a protein exists as a collection of interconverting conformations driven by
thermodynamic fluctuations. Motions on the fast time scale allow a protein to sample conformations in the nearby area of its
conformational landscape, while motions on slower time scales give it access to conformations in distal areas of the landscape.''}

It is important to point out that this is the general picture of protein's functioning and not only  IDP's functioning.

This picture matches perfectly with  the ontic-epistemic approach to the description of dynamics of physical and biological systems. The concrete conformational states 
of a protein are typically unapproachable by experimenters.
From the modern viewpoint, the native state, the functional state,  
cannot be identified with the concrete conformational state. It represents an ensemble of conformational states.  Moreover, the native state is a fuzzy notation 
for a family of functional substates and protein's functioning can be represented as the dynamics in the space of these substates. Each of them corresponds to 
a special subensemble of conformational states. {\small ``Emerging evidence indicates that protein landscapes contain conformational substates with dynamic and structural features that
support the designated function of the protein'',}  \cite{PR}. In this paper we shall identify such ``substates'' with the functional states of a protein. 

Thus we have the two levels description of protein's dynamics:
\begin{itemize}
\item conformational states dynamics: fast and stochastic; 
\item functional states dynamics: slower and deterministic
\end{itemize}
(see, e.g.,  Henzler-Wildman et al. \cite{PT} on time scales in protein dynamics).
 
One of the main consequences of the experimental and theoretical studies is that the dynamics of conformational states, in spite of its stochastic nature, has the functional meta-meaning, i.e., 
it is not reduced to thermodynamical fluctuations. This is a delicate point and we shall try to explain it in more detail. Physically there are, of course, thermodynamical fluctuations. 
But proteins were able to shape these fluctuations in the functional way - to perform transitions (through long chains of  rapid conformations) from one functional state to another, 
see Hammes et al. \cite{PR1},  Ramanathan and  Agarwal \cite{PR2},  Ramanathan et al. \cite{PR3}.      
 
 Nuclear magnetic resonance devices can be used to determine {\it ensembles} of  conformational states of proteins. This is not the end of the experimental studies:
X-ray crystallography allows researchers to identify the most populated states along the landscape. In the functional cycles of proteins, ensembles of conformational
states (determining functional sates) are transferred from one to another. Such ensemble transformations are approachable for the experimental inspection. However, the dynamics
of conformational states in subensembles determining functional states and their jumps from one ensemble to another cannot be detected, at least by existing measurement 
devices, cf. again with the situation in quantum physics.

We emphasize that in contrast to quantum physics, in molecular biology there are no doubts in the existence of conformational states in a protein ``as it is'' (although in general they are 
not accessible experimentally). Of course, there 
is the crucial difference between molecular biology and quantum physics. In biology it is possible ``to see'' the spatial structure of a protein and concrete geometric figures can be used 
for labeling of  conformational states.  In quantum physics, the existing measurement devices do not provide the possibility of determination of a similar fine structure.

\section{Automaton model of protein}
\label{AT}
  
As a model of  the protein dynamics, we consider an automaton \cite{Sakarovitch}   
 $\mathfrak{A}=\langle I,S,V,f,g\rangle,$ where  $I$ ($V$)  is input (output) alphabet, 
$S$  is the set of states,  $f : I\times S\to S$  and $g : I\times S\to V$ are function of transition and output, respectively, i.e., 
depending on the iteration of functioning of the automaton:
\begin{equation}
\label{L1}
    s_{k+1}=f(i_k, s_k),\; k\ge 0.
\end{equation}
\begin{equation}
\label{L2}
v_{k}=g(i_k, s_k),\; k\ge 0.
\end{equation}
At the moment $n = 0$ the automaton $\mathfrak{A}$ is at the state $s_0;$ once fed by the input symbol $i_0 \in  I,$ $\mathfrak{A}$ outputs
the symbol $v_0 = g(i_0, s_0) \in O$ and reaches the state $s_1 = f(i_0, s_0) \in S;$ then $\mathfrak{A}$ is fed by the next input symbol $i_1 \in I$ 
and repeats the routine.

The automaton  $\mathfrak{A}$ transfer words with respect to the alphabet $I$  to words with respect to the alphabet $V$. 
This map, defined on the set of $I$-words and taking values in the set of $V$-words and determined with the aid of 
two functions, $f$ and $g,$ i.e., associated with the automaton $\mathfrak{A}$, is denoted as 
$\mathfrak{F}_{\mathfrak{A}}.$

To construct  the automaton model of the protein dynamics, we have to describe its basic parameters: 

\begin{itemize}
\item \emph{States}.  We assume that a protein consists of 
$N$ connected elementary units, e.g.,  molecules, and that each of them has $p$ degrees 
of freedom. Here $p>1$ is a natural number. Each conformational state of a protein is represented as the vector 
\begin{equation}
\label{STSP} 
\vec s = (x_0,\ldots , x_{N-1}),  \; x_i\in  F_p= \{0,1, \ldots, p-1\}.
\end{equation} 
 
\item \emph{Input and  output alphabets}. The input  alphabet represents  {\it the impact of protein's environment,} e.g., 
temperature, ferments, chemical molecules around it. In our mathematical model it is assumed that influences of the 
environment can be represented numerically, as the elements of the set $I.$ The output alphabet  represents the 
results of observations on a protein; these results are represented as elements of the set $V.$ 

\item \emph{Functions}. The function $f$ is determined by the structure of a concrete protein. 
The function  $g$ is selected depending on the way of observation on a protein: the input alphabet (impact of the
environment), the space of states $S,$ and the transition function $f.$ 

\end{itemize}

{\it Thus in our model  protein by itself is represented as an automaton without output. The output represents observations. Their aim is 
to reconstruct protein's parameters.}

We remark that automata with finite state space $S=\{s_1,...,s_m\}$   (and the fixed  initial state $s_0 \in S$) 
are known as the {\it Mealy machines} \cite{Mealy}. It is not clear whether a protein has to be modeled
as a Mealy machine. Conformational states of protein are determined by a variety of physical parameters, including the geometric configurations. The space of such 
parameters can be infinite and even continuous. Of course, mathematically it is easier to proceed with Mealy machines.

\section{Factor-automaton from observed output sequence}

We remark that the space $S$ of possible conformational states of a protein is huge, 
it contains $N_{\rm{conf}}= p^N$ ($N\sim 300$) elements. Therefore to construct an automaton working on such a space 
and completely describing functioning of the concrete protein is practically impossible. Thus 
the existence of such a complete automaton has basically only a theoretical value.  Possible 
observations cannot determine uniquely protein's state, one fixed  among $N_{\rm{conf}}$ states of $S,$ 
at the instant of measurement, i.e.,   observations are not able to distinguish states from $S.$ 

In reality from observable data we can construct only some {\it factor-automaton} for the ``complete automaton'' $\mathfrak{A}$ 
(ontic automaton).\footnote{This situation reminds emergence of the quantum model, treated as an operational observation model, 
from prequantum models; generation of epistemic state description from the ontic state description, see 
the introduction.}
To construct the factor-automaton representing observations, we consider the following equivalence relation $\rho$ on the
set $S.$  The states $s_1, s_2 $ are $\rho$-equivalent, non-distinguishable with respect to observations, if 
$$
g(i, s_1)=g(i, s_2)\; \forall i\in I.
$$

The factor-set  $S_{\rho}$  of $S$ with respect to $\rho$
is considered as the space of states of the new automaton $\mathfrak{A}_{\rho}.$ The functions $f$ and $g$ 
have to ``respect'' the equivalence relation $\rho,$ i.e., if $$s_1\rho s_2,$$ 
then $$f_{\rho}(i, s_1)=f_{\rho}(i, s_2),\; g_{\rho}(i, s_1)=g_{\rho}(i, s_2), i \in I.$$ 
The  $\mathfrak{A}_{\rho}=\langle I,S_{\rho},V,f_{\rho},g_{\rho}\rangle$ 
is called the factor-automaton of the original automaton $\mathfrak{A}.$  

Thus at the level of observation conformational states of a protein have to be unified in some classes, cf. again 
with ontic-epistemic approach to quantum theory. 

The same model can be used in more concrete biological framework (concertizing  the above general observational
model). We can consider a protein as well specialized biological machine which performs a few concrete functions. In terms
of automaton's states this viewpoint means that on the space of conformational states $S$  it is possible to define some 
equivalence relation which is determined by protein functions. Such a {\it functionality-factorization} also leads to 
the factor-automaton of the original protein-automaton. Thus the states of the factor-automaton can be treated as 
{\it functional states} of a protein. 

Fig. 1  illustrates construction of factor-automata. At this figure the ovals  $C_1,\;C_2,\;C_3,\;C_4,\;C_5$
 represent classes of equivalence with respect to an equivalence relation defined on the space of all possible 
states $S;$ in section \ref{examples} we shall present detailed construction of the concrete factor-automaton.   

\begin{picture}(190,180)(-50,0)
\put(130,130){functional} 
\put(130,120){state} 
\put(120,130){\vector(-1,0){20}}
\put(80,135){\line(0,0){15}} 
\put(80,130){\oval(20,8)}
\put(81,131){\oval(20,8)}
\put(81,130){\oval(20,8)}
\put(80,131){\oval(20,8)}
\put(55,105){\line(1,1){20}} 
\put(105,105){\line(-1,1){20}}
\put(55,100){\oval(20,8)}
\put(56,101){\oval(20,8)}
\put(56,100){\oval(20,8)}
\put(55,101){\oval(20,8)}
\put(105,100){\oval(20,8)}
\put(106,101){\oval(20,8)}
\put(105,101){\oval(20,8)}
\put(106,100){\oval(20,8)}
\put(25,70){\oval(20,8)}
\put(26,71){\oval(20,8)}
\put(25,71){\oval(20,8)}
\put(26,70){\oval(20,8)}
\put(25,75){\line(1,1){21}} 
\put(80,70){\oval(20,8)}
\put(81,71){\oval(20,8)}
\put(80,71){\oval(20,8)}
\put(81,70){\oval(20,8)}
\put(76,75){\line(-1,1){20}}
\put(51,40){\line(-1,1){25}}
\put(53,40){\line(1,1){25}} 
\put(52,25){\line(0,1){15}} 
\multiput(52,10)(0,5){3}%
{\circle*{2}}
\multiput(80,155)(0,5){3}%
{\circle*{2}}
\put(55,127){$C_1$} 
\put(30,97){$C_2$} 
\put(83,97){$C_3$} 
\put(57,67){$C_4$} 
\put(02,67){$C_5$} 

\put(1,0){\small{Figure 1. Functional states of a protein}} 
\end{picture}

$   $

$   $

$   $
 
In future the set of states of the automaton $\mathfrak{A}$ belonging to some equivalence class with respect to 
aforementioned equivalence relation we shall call a generalized state of the automaton. This terminology 
is useful to have the possibility to consider states of the factor-atomaton  $\mathfrak{A}_{\rho}$ 
as sets. In another terminology generalized states are basins, i.e., a basin can be considered as 
a class of equivalence with respect to some equivalence relation $\rho.$

\section{ Example of construction of factor automaton}
\label{examples}

Let us consider automaton $\mathcal{A}=\langle I,S,Y,f,F\rangle,$ 
where  $I=\{0,1,2\}$ denotes the input alphabet, $Y=\{a, b, c, d\}$ denotes the 
output alphabet, and  
$S=\{s_1,s_2,s_3,s_4,s_5,s_6,s_7,s_8\}$  is the state space.  The transition 
function $f$ and the output function $g$ are given by Table 1.

\begin{table}[h]
\centering
\begin{tabular}{|c|c|c|c|c|c|c|c|c|}
\hline
$f(i,s)$& \multicolumn{8}{c|}{Ñîñòîÿíèÿ (s)} \\ 
\hline
Âõîä (i)& $s_1$ & $s_2$ & $s_3$ & $s_4$ & $s_5$ & $s_6$ & $s_7$ & $s_8$\\
\hline
0       & $s_1$ & $s_3$ & $s_1$ & $s_4$ & $s_4$ & $s_7$ & $s_8$ & $s_6$\\ 
\hline
1       & $s_3$ & $s_1$ & $s_2$ & $s_4$ & $s_2$ & $s_5$ & $s_5$ & $s_5$\\
\hline
2       & $s_4$ & $s_4$ & $s_4$ & $s_4$ & $s_7$ & $s_8$ & $s_6$ & $s_8$\\ 
\hline
$g(i,s)$& \multicolumn{8}{c|}{Ñîñòîÿíèÿ (s)}\\
\hline
Âõîä (i)& $s_1$ & $s_2$ & $s_3$ & $s_4$ & $s_5$ & $s_6$ & $s_7$ & $s_8$\\
\hline
0       & $a$ & $a$ & $a$ & $b$ & $c$ & $d$ & $d$ & $d$\\ 
\hline
1       & $a$ & $a$ & $a$ & $b$ & $c$ & $d$ & $d$ & $d$\\
\hline
2       & $a$ & $a$ & $a$ & $b$ & $c$ & $d$ & $d$ & $d$\\ 
\hline
\end{tabular}
\end{table}
\small{Table 1. Transition and output functions for the original automaton.}
\medskip

Table 1 shows that, for this automaton, output depends only on the state. Such automata are known as the {\it Moore machines,} see section \ref{MM}.
The diagram of transitions of the (Moore) automaton determined by Table 1 is represented on the graph of Fig. 2. We repeat that here $g: S \to V,$ i.e., it is independent of the 
input symbol $i.$ 
 $ $

\xymatrix
{s_1\ar@/_1pc/[rrdd]_{1} \ar@/_2pc/[rrddd]_{2} \ar@(l,u)[]^{0}& & & & s_2\ar@/^1pc/[lldd]_{0} \ar@/_2pc/[llll]_{1} \ar@/^2pc/[llddd]_{2}\\
           &  & & &   \\   
           &  & s_3 \ar@/_1pc/[lluu]_{0} \ar@/^1pc/[rruu]^{1} \ar[d]^{2}& &   \\
           &  & s_4 \ar@(d,l)[]^{\{0,1,2\}} & &   \\   
           &  & s_5 \ar@/_1pc/[u]_{0}  \ar@/_3pc/[rruuuu]_{1}  \ar@/_4pc/[llddd]_{2}& &   \\   
           &  & s_6 \ar[u]_{1} \ar@/_1pc/[lldd]_{0}  \ar@/^1pc/[rrdd]^{2} & &   \\
           &  & & &   \\    
          s_7 \ar@/_1pc/[rruu]_{2}  \ar@/^2pc/[rruuu]^{1} \ar@/_2pc/[rrrr]_{0} & & & & 
          s_8 \ar@/^1pc/[lluu]_{0}  \ar@/_2pc/[lluuu]_{1} \ar@(r,d)[]^{2} \\   
}
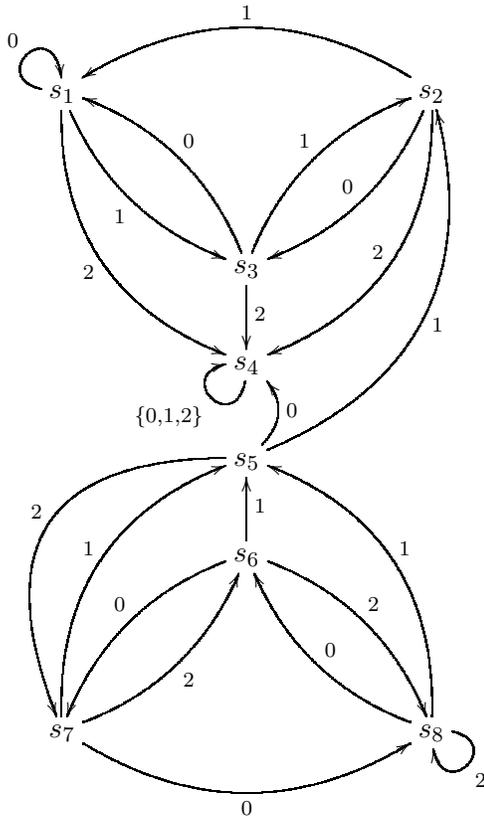
\small{Figure 2. Transition diagram of the original automaton.}
\medskip

We consider the following equivalence relation $\rho$ on the set of states. It is determined with the aid 
of the disjoint partition of the set of states:  
 $S=\{s_1,s_2,s_3\}\cup\{s_4\}\cup\{s_5\}\cup\{s_6,s_7,s_8\}.$ 

The equivalence relation  $\rho$  is consistent with the transition function $f,$ see Table 1,
 i.e. if $s_{\alpha}\rho s_{\beta},$ then 
$f(i,s_{\alpha})\rho f(i,s_{\beta}),\; i\in I;$ $\rho$  is also consistent with 
the output function $g.$

Now consider the new automaton 
$\mathcal{A}/\rho=\langle I,S/\rho,Y,f_{\rho}, g_{\rho}\rangle,$ 
factor-automaton with respect to $\rho,$ which is consistent with the transition and output '
functions (congruence of the original automaton), where   
$$
S/\rho=\{S_1=\{s_1,s_2,s_3\}, S_2=\{s_4\}, S_3=\{s_5\}, S_4=\{s_6,s_7,s_8\}\},
$$
and the transition function  $f_{\rho}$  and the output function 
$g_{\rho}$  are given by Table 2.  

\begin{table}[h]
\centering
\begin{tabular}{|c|c|c|c|c|}
\hline
$f_{\rho}$ & \multicolumn{4}{c|}{Ñîñòîÿíèÿ ($S$)}\\ 
\hline
Âõîä (i)       & $S_1$ & $S_2$ & $S_3$ & $S_4$\\
\hline
0          & $S_1$ & $S_2$ & $S_2$ & $S_4$ \\ 
\hline
1          & $S_1$ & $S_2$ & $S_1$ & $S_3$ \\
\hline
2          & $S_4$ & $S_2$ & $S_4$ & $S_4$ \\ 
\hline
$g_{\rho}$ & \multicolumn{4}{c|}{Ñîñòîÿíèÿ ($S$)}\\
\hline
Âõîä (i)       & $S_1$ & $S_2$ & $S_3$ & $S_4$\\
\hline
0    & $a$ & $b$ & $c$ & $d$\\ 
\hline
1    & $a$ & $b$ & $c$ & $d$\\
\hline
2    & $a$ & $b$ & $c$ & $d$\\ 
\hline
\end{tabular}
\small{Table 2. Transition and output functions for the factor automaton.}
\end{table}
\medskip

The diagram of transitions of the automaton 
$\mathcal{A}/\rho$ determined by Table 2  is presented at Fig. 3.


\xymatrix{& S_4\ar@/_1pc/[d]_{1} \ar@(r,u)[]_{0,2} & \\
          & S_3\ar@/_1pc/[u]_{2} \ar[ld]^{1} \ar[rd]_{0}  & \\
S_1 \ar[rr]_{2} \ar@(l,d)[]_{0,1}   &   &  S_2 \ar@(r,d)[]^{0,1,2} \\
          }
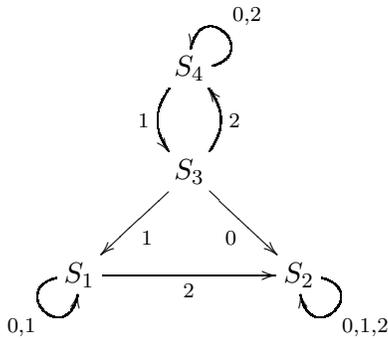
\small{Figure 3. Transition diagram of the factor automaton.}

\section{Protein as a Moore machine?}
\label{MM}

As was remarked in the previous section,  the automaton determined by Table 1 and Fig. 2 belongs to the class of automata known as the {
\it Moore machines} \cite{Moore}:  a finite-state machine (which initial state is fixed) whose output values are determined solely by its current state. Such automata are widely used in computer science and digital electronics.

A bio-system modeled by a Moore machine used environmental inputs only to modify its internal states. Functional outputs are determined solely by the latter.  Such machines, 
although perfectly adaptive to signals from the surrounding environment, have some degree of independence. First they handle environment's signal to modify the state and only later 
use this state to produce a functional output. Thus the information from outside is used not straightforwardly. We can say that it is ``analyzed'' inside the bio-system and only then used 
for the corresponding functional response.  This is the reason to use the Moore machines to model protein's behavior. 

We remark that many digital electronic systems work as Moore machines of the special type, so called {\it clocked sequential systems.}
In such a system the state changes only when the global clock signal changes. This change of the state induces change of the output.  
Clocked sequential systems are one way to solve metastability problems.

We stress that a Moore machine is the finite state machine. However, as was already pointed out in section \ref{AT}, we have reasons both in favor and against using the finite state automata
for modeling protein's behavior. Therefore it is useful to consider so to say {\it infinite state Moore machines.}
 For such an automaton, its state space $S$ can be infinite and even continuous and the output function depend only on the state 

The coupling of our protein model with computer science and digital electronics is the interesting topic for further speculations on the computational structure of functioning 
of cell's components and cells in general. Another important topic is comparison of our modeling of  protein's functioning with mathematical modeling of 
intelligence, both biological and artificial.  The basic model of 
artificial intelligence (or to be more rigorous computing) is given by the {\it Turing machine.}  The latter is definitely more powerful than finite state automata, the Mealy machines
and, in particular, the Moore machines. However, in our modeling of activity of a protein molecular we do not restrict consideration to the finite state automata. Comparison of behavioral 
features of bio-systems modeled with the aid of Turing machines and infinite automata is the interesting problem related to intelligence of bio-systems. In particular, it can be the starting point 
in development of theory of intelligence of proteins.

\section {Problems related to analysis of observational data } 
\label{PRB}

As was pointed out, the automaton model of a protein is an automaton without output. However, initially 
we do not know the structure of this automaton, i.e., the space of states, transition functions, and very often 
neither the input alphabet which encodes influence of environment on behavior of a protein.  To reconstruct 
the parameters of the automaton model of a protein, we have to use the results of observations on protein's 
behaviour. For this aim, the automaton model of a protein is completed  by the output alphabet 
 $V$  and the output function $g$ which are determined by possibilities of observations.
 
 Thus we consider an automaton with output
$\mathfrak{A}=\langle I,S,V,f,g\rangle$, where 
$V$  is the output alphabet and  $g\:I\times S\to V$  is the output function. 
As the result of functioning of this automaton with output we observe a sequence of the form: 
\begin{equation}
\label{OUT}
    v_{k}=g(i_k, s_k),\; k\ge 0.
\end{equation}
We remark that by itself the automaton model of a protein is considered  as the
automaton  $\langle I,S,f\rangle$, 
where, as we shall consider in future, the set of states  $S$ consists of generalized states. 

From observation on the output sequence  $v_{k},\; k\ge 0$ of the automaton  
$\mathfrak{A}$  we can make conclusions about behavior of the protein-automaton as well as about 
its structure (states, transition functions, input sequences ans so on). In particular, we can consider the following 
problems: 
\begin{itemize}
\item  if the transition function  $f_{\rho}$  is only partially defined, then from observations  
$v_{k},\; k\ge 0,$ it can be reconstructed;
\item  it is possible to reconstruct (determine) the input alphabet of an automaton, i.e., to check the influence
of an input symbol on automaton's behavior;  
\item to check the presence of equivalent states by using the output sequence; in other words, 
to clarify the structure of the equivalence relation on the set of states, i.e., states of the form
   $\vec s = (s_0,\ldots ,s_{N-1})$;
\item to check matching of an automaton with the output data (observed in experiments);
\item to determine the initial state of the automaton $\mathfrak{A}$.
\end{itemize}

Since the automaton $\mathfrak{A}$ can have generalized states, its output sequences  $v_{k},\; k\ge 0$ 
can have the following specific feature. An output sequence of the automaton $\mathfrak{A}$ 
can contain a long sequence of equal elements. Such sequences appear if the automaton is being 
in some fixed generalized state. If the output sequence is considered as realizations of some random 
variable $\gamma,$ then the presence of long series can imply some ``pathological features'' 
of $\gamma,$ e.g., $\gamma$ does not have the mathematical expectation. Such features make 
difficult construction of a proper statistical criteria for solution of aforementioned problems.

\medskip

To illustrate  this situation, we consider behavior of the automaton  $\mathfrak{A}$ 
in a neighborhood  of a generalized state $C$ as it is shown at Fig. 4.

\begin{picture}(160,80)(-85,120)


\put(110,140){\oval(50,30)[tr]}
\put(110,140){\oval(50,30)[br]}

\put(80,140){\oval(50,30)[tl]}
\put(80,140){\oval(50,30)[bl]}

\multiput(90,125)(5,0){3}%
{\circle*{2}}

\multiput(90,155)(5,0){3}%
{\circle*{2}}

\put(55,140){\circle*{4}}
\put(58,150){\circle{4}}
\put(70,155){\circle{4}}

\put(135,140){\circle{4}}
\put(132,150){\circle{4}}
\put(123,155){\circle{4}}

\put(132,130){\circle*{4}}
\put(123,125){\circle*{4}}

\put(58,130){\circle*{4}}
\put(70,125){\circle*{4}}
\put(40,150){$i_1$}
\put(60,140){$s_1$}
\put(120,140){$s_2$}
\put(145,150){$i_2$}
\put(90,165){$C$}
\put(5,140){$C_{-1}$}
\put(170,140){$C_{1}$}

\qbezier(33,142)(43,147)(53,142)
\put(33,142) {\vector (-2,-1){3}} 
\put(28,140){\circle*{4}}
 \qbezier(30,138)(40,133)(50,138)
\put(51,138) {\vector (2,1){3}} 

\qbezier(140,142)(150,147)(160,142)
\put(140,142) {\vector (-2,-1){3}} 
\put(163,140){\circle*{4}}
 \qbezier(137,138)(147,133)(157,138)
\put(158,138) {\vector (2,1){3}} 

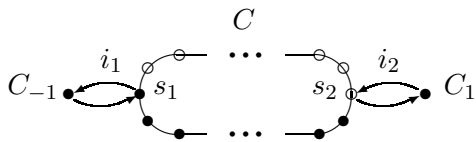
\put(1,100){\small{Figure 4. Unconventional probabilistic behavior.}}

\end{picture}

$   $

$   $

$   $

There are three generalized states $C_{-1},\;C,\;C_1$. 
Each of generalized states $C_{-1},\;C_1$ contains just one  conformational state. 
The generalized state  $C$  consists of $2p^r,$ where $p>1$  is a natural number,
conformational states (the states  $s_1,..., s_2$ belong to 
 $C$).  Conformation states composing  $C$  form two special classes,
at Fig. 4 they are represented by white and black spots. Each class contains 
 $p^r$  conformational states. 

The automaton is functioning in the following way.  If it is in state  $C_{-1}$ ($C_{1}$), 
then, for each input, the automaton is transferred into the conformational  state $s_1$ ($s_2$) 
belonging to the generalized state $C.$ 
If the automaton is in one of conformational states composing $C,$ 
excluding the states  $s_1$ and $s_2,$  and marked by white (black) points,  then for 
each input the automaton is transferred into the closest from the left-hand (right-hand) side state. 
Thus in the generalized state $C$ the automaton goes through conformational states  composing  $C$ 
cyclically. For the states $s_1$  and $s_2$,  its behavior is more complex. There are two possible transitions 
scenarios. If input symbol is  $i_1$ ($i_2$),
then the automaton jumps into the generalized state 
 $C_{-1}$ ($C_1$).  If the input symbol is different from 
$i_1$ ($i_2$), then the automaton stays in the generalized state $C$  and its state is changed in accordance 
with the general law of cyclic dynamics insied $C.$

Suppose now that the output function of the automaton $g$ 
depends only on the generalized state, i.e., it does not depend on input and automaton's output 
$v_{-1}$ is observed if the automaton  is in the state $C_{-1}$ (respectively, $v$ and $v_1$ for $C$,  $C_{1}$ ).

Since in the state  $C$ the automaton can stay   $\xi p^r$  steps of dynamics, where 
$\xi$  is some integer valued random variable, in general the relative frequency of appearance of the symbol 
$v$  in the output sequence can fluctuate, without stabilization to any constant - probability.   Thus the probability 
$P(C)$ of the state $C$  is not well defined. Besides of this 
``probabilistic pathology'' (violation of von Mises principle of stabilization of relative frequences),  we point to 
another feature of behavior of relative frequencies: for the states $v_{-1}$ and $v_1,$ they approach zero 
(when the length of the output sequence approaches infinity). Thus the probabilities of appearance of this automaton in the 
states   $C_{-1}$,  $C_{1}$ equal to zero. Therefore from the probabilistic viewpoint these states can be simply 
eliminated from the model. However, this will lead to a biologically inadequate model. 

For such sequences of observations, we can use the apparatus of generalized probability theory based on representation 
of probabilities not by real numbers from the segment $[0,1],$ but by so called $p$-adic numbers, where $p>1$ 
is a prime number. In this framework the relative frequencies (which are always rational numbers) are embedded
into the fields of $p$-adic numbers $\mathbf Q_p$ and their limits are considered in these fields. In this way
the probabilities $P(C)$ and $P(C_{-1}),  P(C_{1})$  are well defined and the last two of them are nonzero. In this paper 
we cannot go in more details, see \cite{KHR_P1},  \cite{KHR_P2}.

\section{Automaton model: some features of protein's behavior and interpretations}
\label{AMI}

In this section we shall consider some special features of protein's behavior implied by our automaton-model
and possible interpretations of these features.  

\begin {enumerate} 
\item {\it Testing the presence of generalized states.}  Consider the following situation represented at Fig. 5.1 and 5.2. 
In the first case the automanton has only two generalized states for which we observe outputs  
$v_{-1}$ and $v_1,$ respectively. In the second case (Fig. 5.2)  there is the generalized state $C$ between the 
states $C_{-1}$ and $C_1.$  However, when this automaton is in the state  $C$
no outputs are observed.  If the automaton is in the state $C_1$ ($C_{-1}$), 
but has not yet moved to the state  $C_{-1}$ ($C_1$), 
then the symbol  $v_1$ ($v_{-1}$) is observed at the output sequence .
 
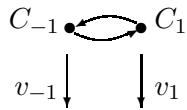
\begin{picture}(80,80)(0,100)

\put(55,140){\circle*{4}}

\put(5,140){$C_{-1}$}
\put(60,140){$C_{1}$}

\qbezier(33,142)(43,147)(53,142)
\put(33,142) {\vector (-2,-1){3}} 
\put(28,140){\circle*{4}}
 \qbezier(30,138)(40,133)(50,138)
\put(51,138) {\vector (2,1){3}} 

\put(27,130) {\vector (0,-1){20}}
\put(55,130) {\vector (0,-1){20}}
\put(7,115){$v_{-1}$}
\put(60,115){$v_{1}$}

\put(1,40){\small{Figure 5.1. No intermediate state}}
\end{picture}

\begin{picture}(160,80)(-100,20)
\put(110,140){\oval(50,30)[tr]}
\put(110,140){\oval(50,30)[br]}

\put(80,140){\oval(50,30)[tl]}
\put(80,140){\oval(50,30)[bl]}

\multiput(90,125)(5,0){3}%
{\circle*{2}}

\multiput(90,155)(5,0){3}%
{\circle*{2}}

\put(55,140){\circle*{4}}
\put(135,140){\circle*{4}}

\put(90,165){$C$}
\put(5,140){$C_{-1}$}
\put(170,140){$C_{1}$}

\qbezier(33,142)(43,147)(53,142)
\put(33,142) {\vector (-2,-1) {3}} 
\put(28,140){\circle*{4}}
 \qbezier(30,138)(40,133)(50,138)
\put(51,138) {\vector (2,1) {3}} 

\qbezier(140,142)(150,147)(160,142)
\put(140,142) {\vector (-2,-1) {3}} 
\put(163,140){\circle*{4}}
 \qbezier(137,138)(147,133)(157,138)
\put(158,138) {\vector (2,1) {3}} 

\put(60,140){$s_1$}
\put(120,140){$s_2$}

\put(27,130) {\vector (0,-1) {20}}
\put(165,130) {\vector (0,-1) {20}}
\put(32,115){$v_{-1}$}
\put(172,115){$v_{1}$}

\put(100,40){\small{Figure 5.2. Intermediate state}}
\end{picture}

$   $

The output sequence can be used to determine the presence of an intermediate state $C$ which can be present 
in a protein. The presence of such a state $C$ implies that the output sequence  of the 
automaton presented at Fig. 5.2 contains series of elements  $v_{-1}$ and $v_{1}.$ The lengths of series can be considered 
as values of some random variable. In the case of the automaton presented at Fig. 5.1  the output sequence 
also consists of the elements  $v_{-1}$ and $v_{1}.$ In principle this sequence also can contain long series  of  
these elements, but their probabilistic structure differs from the probabilistic structure of the series in the output 
sequence generated by the automaton of Fig. 5.2.  We now point to one of the important marks of the presence of an intermediate 
state $C:$  in the sequence produced by the automaton of Fig. 5.2,  the law of 
stabilization of relative frequencies of appearance of  $v_{-1}$ and $v_{1}$ can be violated. As was remarked, 
for analysis of such statistical data, $p$-adic probability methods can be applied. 

\item {\it The presence of generalized states and protein's transition time.} Our model with generalized states as composed 
of internal (conformational) states of proteins can explain the well known experimental fact that transition of a
protein from one functional state to another can take long time and, moreover, that the interval of transition time
can vary essentially from transition to transition, Henzler-Wildman et al. \cite{PT}.   Thus the descriptive level based on ontic states of proteins 
has some explanatory power for observable (epistemic) features of proteins.  

\item {\it Hierarchy.} By using the automaton model of a protein the set of generalized states, functional states of 
proteins,  can be endowed with hierarchic structures, in particular, tree-like. We remark that the feature of the 
model that  transition from one functional state to another is performed  through other generalized states (Ramanathan et al. \cite{PR}) can
 be explained even without appealing to a hierarchic structure in proteins. (For example, the graph 
represented at Fig. 6 is, in fact, the binary tree. However, the presence of the tree-like structure is not straightforwardly 
visible at Fig. 6.)    To fill hierarchy with nontrivial content (i.e., not just the graphic scheme), perhaps we have to 
use some additional criteria. For example, a hierarchy determining rule can be based on the magnitude of 
the total energy of transition  from one functional state to another or the magnitude of the potential energy barrier 
separating these states, cf. with works \cite{}.

\begin{picture}(80,50)(-95,0)

\put(0,45){\circle*{4}}
\put(0,10){\circle*{4}}
\put(0,10) {\line (1,1) {17}}
\put(0,45) {\line (1,-1) {17}}
\put(18,27){\circle*{4}}
\put(18,27) {\line (1,0) {17}} 
\put(35,27){\circle*{4}}
\put(35,27) {\line (1,0) {17}} 
\put(52,27) {\line (1,1) {17}}
\put(52,27) {\line (1,-1) {17}}
\put(52,27){\circle*{4}}
\put(70,45){\circle*{4}}
\put(70,10){\circle*{4}}

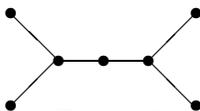
\put(17,0){\small{Figure 6. Tree-like structure of protein-automaton.}}
\end{picture}

\item  {\it Functional inseparability.} In the framework of the protein-automaton  we can consider the following 
extremal case. For example, the factor-automaton consists of a single generalized state  $C$ -  the native state of a protein. In such a situation
we are not able to separate functional abilities of conformational states.  Even by knowing that the number of conformational states is huge 
we have the trivial factor-automaton. Moreover, the impossibility of epistemic splitting of   $C$ strongly supports the protein  
structure-function paradigm.  It is very attractive to extend this uniqueness  from the epistemic  level to the ontic level.  And this happened 
at the first stage of development of molecular biology (in fact, the very long stage), see section \ref{MB}. 
  
However, from the modern viewpoint \cite{PR}, the space of conformational (ontic) states has to be 
divided into classes - generalized  (epistemic) states. The latter states have to be distinguishable 
from the functional viewpoint (as representing observable behavior of a protein). The behavior of a protein
inside a generalized state can be considered  as ``random and   equiprobabilistic''
and having no influence on functioning of a protein.  

We remark that from the  protein-automaton viewpoint  the rather common thesis that 
proteins are biological machines  posses the following constraint on
the relative number of generalized states $N_{\rm{gen}}.$ 
It has to be small, but at the same time not too small. On one hand, if $N_{\rm{gen}}$ is large,  then such a biological machine 
has to perform a large number of various functions or if all these generalized states (``substates of the native state''  in terminology of  Ramanathan et al. \cite{PR}) are involved in a single 
functional cycle, then its performance will take too much time.   This contradict to the experiment and the thesis 
that a protein is similar to a machine. On the other hand, if $N_{\rm{gen}}$ is very small  (e.g., just one as 
in the above example), then behavior of such a biological system is trivial (from the epistemological viewpoint).

\end {enumerate} 

\section{Mathematical formalism}
\label{MFOR}

In the previous sections we tried to minimize the use of the mathematical apparatus; in fact, we proceed with only one mathematical structure, automaton, 
representing states' dynamics. Now we want to be more precise in mathematical terms. 

\subsection{$P$-adic model of the space of conformational states}
\label{MM}

For a mathematical model of the state space of the protein-automaton, we can consider infinite sequences (cf. with (\ref{STSP})):  
\begin{equation}
\label{STSPI} 
\vec s = (x_0,\ldots , x_{N-1},...),  \; x_i\in F_p= \{0,1, \ldots, p-1\}.
\end{equation} 
Here $p>1$ is a natural number. 
Denote the space of such sequences by the symbol $\mathbf{Z}_p.$ For mathematical modeling, it is useful to endow $\mathbf{Z}_p.$ with an algebraic structure similar 
to the algebraic structure on real numbers. Each sequence is represented as the series of the form: 
\begin{equation}
\label{STSPI1} 
\vec s = \sum_{i=0} ^\infty x_i p^i, \; x_i\in F_p.
\end{equation} 
We remark that the set of natural numbers $\mathbf{N}$  is the subset of  $\mathbf{Z}_p$ consisting of finite sums (expansions of natural numbers with respect to the base $p).$ 
Moreover, negative integers can be also represented in the form of such power series, by using so called complement code representation which is standard for computer science.
 For example, if $p=2,$ then  $-1= 11111....1..... = \sum_{i=0}^\infty 2^i.$ Thus even the set of integers $\mathbf{Z}$ can be embedded in $\mathbf{Z}_p.$ The operations of addition, subtraction and multiplication
 can be extended from $\mathbf{Z}$ onto $\mathbf{Z}_p.$ We obtain the algebraic structure called a {\it ring.} Division is in general not well defined, see, e.g., \cite{KHR_P2} for details. 

The space $S=\mathbf{Z}_p$ can serve as the mathematical model of the (ontic) state space of the protein-automaton. It can be endowed with a metric which is so called ultrametric, see section \ref{UM}.
Thus {\it the (ontic) state space of the protein-automaton is the ultrametric space.}  Ultrametricity expresses mathematically the hierarchic structure of the state space, see section \ref{UM}.
This is the simplest ultrametric mathematical model.  

In a more general model the ranges of digits $x_i$  need not coincide, i.e., $x_i \in F_{p_i}=\{0,1,..., p_i-1\},$ where $p_i>1$ is a natural number.
The latter model is more natural, since encoding unit $i$ in the digital representation of protein's state, see  (\ref{STSP}) and (\ref{STSPI}), can have its own degree of  complexity reflected in the number   $p_i.$ 
We set ${\bf p} =(p_0,...,p_k,...).$ Denote the space of sequences  with $x_i \in  F_{p_i}$ by the symbol $\mathbf{Z}_{{\bf p}}.$ This space can also be endowed with 
the algebraic structure and the ultrametric. Unfortunately, analysis on $\mathbf{Z}_{{\bf p}}$ is very complicated. In particular, nothing was done to develop theory of dynamical systems 
for such spaces. And the latter can serve as a tool to construct concrete examples of protein's state dynamics, see section \ref{UM}.

\subsection{Ultrametricity}
\label{UM}

Typically a hierarchic structure in a physical or biological system can be mathematically presented as ultrametricity, i.e., the {\it ultrametric space} structure 
on the state space. The ultrametric model is widely used in theory of complex disordered systems (e.g., spin glasses) \cite{K1}-\cite{K3}, in genetics \cite{D1}- \cite{D2},
cognition and psychology \cite{KHR0}-\cite{GI2}.     
We recall briefly the definition of an ultrametric space and the main features of its system of neighborhoods given by 
ultrametric  balls. We start with the terminological remark: ultrametric spaces are also known as    {\it non-Archimedean spaces} \cite{KHR_P2}.  

Let $S$ be a set (in our case the state space of a protein, the space of its conformational states). Let $d: S \times S \to \mathbf R_+$ be a metric (distance). It is called an {\it ultrametric}
if, besides the standard triangle inequality, it satisfies the  {\it strong triangle inequality:} 
\begin{equation}  
\label{STE}
d(x,y) \leq \max[d(x,z), d(z,y)], x,y,z \in S.
\end{equation}
This inequality for a metric implies many unusual features of the system of neighborhoods  in $S$ given by the balls 
$$
U_r(a)=\{x \in S: d(x,y) \leq r \}, U_r^-(a)=\{x \in S: d(x,y) <r \},
$$
where $r>0$ and $a \in S.$ 
\begin{enumerate}
\item All balls are topologically open and closed at the same time - ``clopen''.  
\item They are either disjoint or one is subset of another. 
\item Any point $b$ belonging 
to the ball can be chosen as its center. Thus if $b \in U_r(a),$ then $U_r(a)=U_r(b).$ The same holds for balls $U_r^-(a).$  We can say that ``all points of a ball-neighborhood have 
equal rights.''
\end{enumerate}

We now turn to our factor-automaton model of protein functioning. If we take any conformational state $s$ belonging to a functional state $C,$ then $s$ completely determines 
the latter - as a representative of the factor class $C.$ Consider the functional state $C$  as {\it a neighborhood of the conformational state} $s.$ Then any $s \in C$ can 
be treated as the center of this neighborhood. Thus all conformational states belonging to a functional state ``have equal rights'' (from the observational viewpoint they produce 
the same output).  This is an exhibition of ultrametricity of the state space of the automaton-protein.  This is so to say epistemic ultrametricity, it is generated by functional factorization 
of conformational states. By using ultrametric spaces we can represent functional states by balls $C=U_r(s).$    

Consider the mathematical model in which  the space of conformational states $S=\mathbf Z_p$ (the ring of $p$-adic integers) and  all functional states are given by 
balls of the same radius $r=1/p^k.$ The set of all conformational states $S$ can be represented 
as the disjoint union of the balls of the radius $r: S=\cup_{j=0}^{p^k-1} U_r(j).$  This state space of balls can be obtained as the output of the factorization procedure on $S$ 
based on the $\rm{mod} \; p^n$ equivalence relation: for $x,y \in  \mathbf Z_p, x \sim y$ if and only if $x-y$ is divisible by $p^n$ (see (\ref{STSPI1}) for number-theoretic 
presentation of elements of  $\mathbf Z_p).$      

Functioning of the protein-automaton can be described as a $p$-adic dynamical system
\cite{A1}-\cite{A3}, see appendix 2. 

\section{Conclusion}

We proposed to model protein behavior by using theory of automata. 

We pointed out to similarity between modeling of behavior of proteins and quantum systems.  In analysis of quantum foundations some authors explore 
{\it the ontic-epistemic methodology}, see Appendix 1. Quantum mechanics is treated as the 
epistemic (observational) model which can be emergent from a causal ontic (subquantum) model, see, e.g., Atmanspacher et al. \cite{H1}-\cite{H4},  
Khrennikov \cite{KHR_PRS1, KHR_PRS2},  't Hooft \cite{TH1} and Anashin \cite{ANew}, \cite{ANew1}.\footnote{At the same time other experts follow the tradition 
of the Copenhagen school and reject the possibility of  emergent treatment of quantum mechanics, see the work of Plotnitsky and Khrennikov for detailed 
analysis \cite{PL_KHR}.} The states of the latter are not directly approachable for external inspection.
In some models \cite{H1}, \cite{KHR_PRS1, KHR_PRS2},  \cite{TH1}  transition from the subquantum (ontic) level  to the quantum (observational) level is performed through 
factorization of the space of subquantum states. Thus the states described by the quantum formalism are treated as the symbolic images of factor classes of ontic states.  
Moreover, in quantum mechanics there are widely used dynamical systems to model  ontic$\mapsto$epistemic transition, see \cite{H1}, \cite{KHR_CSH}.

This is the good place to remark that  recent development of physics is characterized by creation of  a number of 
novel automaton models describing state dynamics. This activity was culminated 
in `t Hooft's model of totally deterministic dynamical universe \cite{TH1} functioning as a cellular automaton.\footnote{We emphasize that one has to distinguish 
sharply theory of cellular automata from general theory of automata used in our paper. The common use of the word ``automaton'' can be misleading. Here we have no possibility 
to go into details. We just point to the reader that this paper {\it is not about cellular automata!}} We also mention the recent work of Anashin \cite{ANew}, \cite{ANew1}, an attempt to derive 
the wave structure of quantum theory from theory of $p$-adic dynamics systems.    

Now we tern to molecular biology. The dimension of the space of possible conformation states of proteins is huge and these states are not approachable for direct external 
inspection. Therefore it is natural to proceed similarly to the ontic-epistemic approach to quantum mechanics  and to work jointly with two coupled
 models of protein's behavior:
\begin{itemize}
\item the ontic model describing dynamics of conformational states;
\item the epistemic model describing functional states of   a protein. 
\end{itemize}
The dynamics of conformational states is described by an automaton, discrete time and state dynamical machine. Input symbols  encode the influences of the environment (the state of the surrounding 
cell and signaling from other cells). The transition function $f=f(i, s)$ describes transitions between internal states of protein (conformational states) generated by interaction of a protein molecular 
with the cell-environment;
the output function $g=g(i,s)$ describes  functional outputs of the protein molecular. The dynamics of functional states is described by the factor-automaton generated through functional 
equivalence relation on the space of conformational states and the corresponding factorization of the state space and the transition and output functions. 
Here it is the good place to cite once again  Ramanathan et al.  \cite{PR}: {\small  ``Motions on the fast time scale allow a protein to sample conformations in the nearby area of its
conformational landscape, while motions on slower time scales give it access to conformations in distal areas of the landscape.''} The mentioned fast scale dynamics is the automaton 
dynamics of ontic states, slow scale dynamics describes transitions of epistemic states. The later can be slow enough comparing with the fact scale transitions, because any epistemic (functional) 
state is composed of a large number of ontic states. The protein molecular can spend a lot of time inside such a cluster of conformational states before to jump to the next cluster representing 
another epistemic state. 

This factor-space structure of the space of functional states can lead to nonclassical probabilistic behavior. We remark that in the complete accordance with the ontic-epistemic approach 
nonclassicality of probability is a purely epistemic feature. At the ontic level, classical probability works well.  Such a model can generate very exotic probability distributions, not only 
quantum-like, cf.  \cite{H1}-\cite{H4},  \cite{KHR_PRS1, KHR_PRS1},  \cite{TH1}, but even so called $p$-adic probability distributions \cite{KHR_P2}. The latter were invented in $p$-adic 
(non-Archimedean) physics to provide the probabilistic interpretation of wave functions valued in the fields of $p$-adic numbers or more general non-Archimedean (ultrametric) number 
fields. Then a few theoretical biological models generating $p$-adic probabilities were invented and theory was completed with numerical simulation. Now we hope that $p$-adic probability 
theory can find real applications in molecular biology. Of course, these are just preliminary conceptual considerations. 
 
In appendix 2 we presented coupling (due to Anashin) of theory of automata with mappings in the fields of $p$-adic numbers. We tried to make the first steps to describe the class 
of automata and corresponding $p$-adic maps matching functioning of proteins. This will be the main direction of our further studies. To proceed in this direction, we have to analyze
functioning of proteins in more detail, especially from the biological side (and not only  dynamical and information processing constraints discussed in appendix 2).

\section*{Acknowledgments} 

This paper was written under support of the grant of the Faculty of Technology of Linnaeus University, Modeling of Complex Hierarchic systems and 
the EU-project "Quantum Information Access and Retrieval Theory" (QUARTZ), Grant No. 721321

\section*{Appendix 1: Quantum-like models in biology}

This paper is not about quantum bio-physics, cf. \cite{Arndt}, \cite{M1}-\cite{M3}. It is closer to  works on so called {\it quantum-like models}
exploring the apparatus of quantum probability and information theories to model biological phenomena \cite{34}--\cite{polina3}, \cite{M2}, 
\cite{M3}. In quantum-like modeling one does not try to study genuine 
quantum physical processes in a bio-system. The latter is considered as a black box processing information in accordance with the laws of quantum probability and information. 
Moreover, quantum probabilistic-informational behavior need not be explicitly coupled to quantum physical processes. Quantum-like behavior can be exhibited by macroscopic 
bio-systems operating at time, space, and temperature scales which do not match the corresponding scales of the quantum physics. 

The most successful applications of 
the quantum-like approach are in modeling brain's functioning, cognition, decision making, psychology, see, e.g., the monographs  \cite{UB}, \cite{11}, \cite{30a}. 
Here the brain is treated as a black box with inputs from environment (both physical and mental) and outputs in the form of perceptions, decisions and judgments. The probabilistic-information 
structure of interrelation between inputs and outputs cannot be described mathematically by the laws of classical probability and information. Therefore alternative (``nonclassical'') theories 
have to be proposed. The quantum formalism is the most well developed, advanced and approved in numerous applications. Therefore it is reasonable to apply this formalism 
as an alternative to the classical probability and information formalisms. This was done in the aforementioned applications outside of physics.  Of course, one cannot guarantee that in biological 
applications nonclassicality coincides with quantumness. It can happen that from the probabilistic and information viewpoints behavior of biological systems is even more complicated  than behavior of quantum systems; may be
the brain can exhibit informational patterns which structure is more complex than the structure of genuine quantum patterns. May be novel mathematical formalisms have to be designed 
for biological purposes. 

Once again we remind that the quantum-like  model  of brain's functioning differs from the ``quantum physical  brain model'' in the spitit of R. Penrose \cite{RP} and S. Hameroof 
\cite{SH}.  The latter tries to connect cognitive actions of the brain at the level of consciousness with physical processes in the brain at the quantum level. The main bio-physical structure 
involved to this process are micro-tubular. It is claimed that they can be used by the brain for quantum information processing. In particular, they can exhibit such a basic feature of quantum 
physical systems as entanglement. Justification of this conjecture is the big challenge. For the moment, we cannot reject completely the possibility that the brain really works in this way by exploring 
quantum physical processes, e.g.,  at the level of micro-tubular. However, to justify this theory of brain's functioning one has to solve many problems and some authors presented very heavy 
arguments that the hot amd macroscopic brain cannot ``enjoy'' exciting features of the quantum world, see, e.g., Tengmark \cite{Tegmark}. However, the creators of the      quantum physical  brain model
argue that quantum features still can be preserved in the brain, at least for time-intervals sufficient to perform a kind of quantum computation leading to cognitive processing of information.
The lovely debate have been continued already 30
 years.  It may take many years to clarify the situation (if such clarification is possible at all in the framework of existing physical 
theory).  
         
One of the main problems in justification of the quantum-like approach to modeling of cognitive behavior of bio-systems is the resolution of the famous {\it  problem of hidden variables.} 
For biologists, we say a few words about this problem. 

The state  $\psi$ of a classical physical system determines the outcomes of all observables which can be measured for this 
state. Observables are represented as functions on the state space,  $\psi \to f(\psi).$  For example, in classical mechanics the state $\psi$ is given by the pair $\psi=(x,p),$ position and 
momentum (the coordinates in phase  space). Here observables can be treated as functions on the phase space, $f=f(x,p). $  In classical statistical mechanics the state $\rho$ is 
represented by  the probability distribution on the phase space; observables are also defined as functions on the phase space. 
This kind of states, statistical states, do not determine the measurement outcomes, but only the probabilities of these outcomes.
Thus the average 
of the observable represented by the function $f$ in the state given by the probability distribution $\rho$  is calculated by the rule of classical probability theory:
\begin{equation}
\label{LCP}
\langle f\rangle_\rho= \int f(x,p) d \rho(x,p) .
\end{equation}
If the probability has the density $\rho(x,p),$ then the average is given by the integral:
$\langle f\rangle_\rho= \int f(x,p)  \rho(x,p) dx dp.$

The state of a quantum system is represented by a (normalized) vector  $\psi$ of complex Hilbert space $H,$ i.e., $\langle \psi\vert \psi \rangle =1,$
where $\langle \phi_1 \vert \phi_2 \rangle$ denotes the scalar product of vectors $\phi_1, \phi_2 \in H.$ 
Quantum observables are represented by Hermitian operators. 
Similarly to the state used in classical statistical mechanics, it determines only the probabilities of 
outcomes. However, in contrast to classical statistical mechanics, the quantum average is represented as 
\begin{equation}
\label{LCP1}
\langle A \rangle_\psi= \langle A \psi \vert \psi\rangle.
\end{equation}

From the very beginning of quantum mechanics, the following question disturbed its fathers: {\it Is it possible to introduce a quantum analog of the classical phase space?} More generally 
this is the question about the possibility of functional representation of quantum observables and measure-theoretic representation of quantum states. In particular, one wonders 
whether it is possible to represent quantum average (\ref{LCP1}) in the classical integral form (\ref{LCP}).

 Since for the moment physical parameters determining the outputs of measurements are unknown
\footnote{In contrast to classical mechanics, the pair position-momentum,  $\psi=(x,p),$ cannot serve as such parameters, because the position and momentum cannot be measurement jointly. 
By the orthodox version of the Copenhagen interpretation they even ``do not exist''.}, they got the name {\it hidden variables.}
 Two orthogonal positions were presented by  Einstein and Bohr, respectively.
Einstein was sure that such variables determining the values of quantum observables exist and this is just a matter of time, soon or latter the formalism with hidden variables reproducing quantum 
measurement theory will be designed. Bohr was sure that this will never happen, hidden variables do not exist. Bohr's position was merely philosophical. Later it was transformed into 
mathematical statements known as ``no-go'' theorems. The first of them was proven by J. von Neumann. However, it was criticized as unphysical. The most famous no-go theorem 
was formulated by J. Bell. It is based on the inequality for correlations known as {\it Bell's inequality.} A theory with hidden variables should produce correlations which satisfy this 
inequality. But the quantum formalism implies correlations violating this inequality. The experiment seems to be in favor of the quantum prediction. We emphasize that Bell's no-go theorem 
differs crucially from von Neumann's  theorem. Bell lifted up the issue of {\it nonlocality} which first mentioned by Einstein. This issue was not present in 
von Neumann's theorem.  The main point in our discussion is that  a nonlocal hidden variable model can match the predictions 
of quantum mechanics. Correlations based on signaling between subsystems can violate Bell's inequality.  In physics such signaling is impossible, because subsystems, e.g., 
two entangled photons, are separated for a long distance such that signaling should have the speed extending the speed of light. Therefore quantum nonlocality appears 
as a mystical physical phenomenon, {\it the spooky action at a distance.} 

Now we turn again to biology. In contrast to quantum physics,  here ``hidden variables do exist''. Moreover, they are not hidden at all. These are states of neurons in brain science, states 
of proteins, genes, cells  in molecular biology and so on. It seems that those who apply the quantum-like approach are in trouble. Quantum theory says that there are no (local) hidden variables,
but biology says that they exist (and not even hidden). The key word here  is (non)local. In biology we need not be afraid of nonlocal models, because they are nonlocal only formally. In reality 
nonlocal contributions to observations can be generated by signaling between subsystems, e.g., between cells. This signaling is not mystical, since its speed does not exceeds the speed of light. 
The crucial point  is that biological systems are negligibly small in size (comparing with light's velocity), cf. \cite{AB}. Thus we can apply the quantum formalism to biological phenomena without to collide 
with no-go theorems! 

As was pointed out, quantum-like modeling was very successful i modeling of cognition and decision making. However, recently quantum operational 
 methods were extended to model behavior of micro-biological systems, e.g.,  cells, genomes, epigenomes,  and proteins \cite{b1}, \cite{SSB1}. Thus we can speak 
 about {\it quantum information biology} as unifying information processing at all scales, from molecular biology to cognition and social science, see \cite{BOOK}, \cite{FOOP} 
for details.

\section*{Appendix 2: $P$-adic mapping representation of epistemic automata} 

In this section following V. Anashin \cite{A4a} we shall consider representation of apparata by mappings of the ring of $p$-adic numbers $\mathbf Z_p.$  
We start with presenting theory of automata in more detail. We remind that the {\it initial automaton} $\mathfrak{A}(s_0)=\langle I,S,V,f,g, s_0\rangle$ is an automaton,  where one state
$s_0 \in S$  is fixed; $s_0$ is called the initial state. In particular, the Mealy machines  \cite{Mealy} (and hence the Moore machines \cite{Moore})
 are initial automata with finite state spaces.

 \begin{center}
\begin{picture}(250,100)
\put(0,50){\vector(1,0){90}}
\put(130,50){\vector(1,0){90}}
\put(0,35){$\cdots,i_k,\cdots ,i_1 ,i_0$}
\put(140,35){$y_0,y_1, \cdots ,y_k,\cdots$}
\put(95,80){$s_0,\cdots $}
\thicklines
\put(90,20){\line(0,1){70}}
\put(90,90){\line(1,0){40}}
\put(90,20){\line(1,0){40}}
\put(130,20){\line(0,1){70}}
\put(140,70){$\mathfrak{A}_{s_0}$}
\put(105,0){{\small Figure 1: Functioning of initial automation}}
\end{picture}
\end{center}

In theory of automata it is often assumed that there exists a state $s_0 \in S$ such that all the
states of the automaton $\mathfrak{A}$ are reachable from $s_0;$ that is, given  state $s  \in S,$ there exists
input word $w$ over alphabet $I$ such that after the word $w$ has been fed to the
automaton $\mathfrak{A}(s_0),$ the automaton reaches the state $s.$ 

To the automaton $\mathfrak{A}$ we put into a correspondence the family  of all  initial automata 
$\mathfrak{A}(s)=\langle I,S,V,f,g, s\rangle, s \in S.$

Now we assume that both alphabets $I$ and $O$ are $p$-element alphabets: $I = O = \{0, 1, . . . , p - 1\} = F_p;$ so further
the word 'automaton' stands for initial automaton with input/output alphabets
$F_p.$ A typical example of an automaton of that sort is the 2-adic {\it adding machine}, 
$\mathfrak{A}(1)=\langle F_2, F_2,F_2, f, g, 1\rangle,$ where $f(i, s) = is (\rm{mod}\;  2),  g(i, s) = i + s (\rm{mod}\;  2)$ 
for $s \in S = F_2, i\in I = F_2.$

Given an automaton $\mathfrak{A}(s)=\langle I,S,V,f,g, s\rangle,$  the automaton transforms
input words  (with respect to the alphabet $F_p)$ of length $n$ into output words of length $n;$ that is, $\mathfrak{A}(s)$ maps
the set $W_n$ of all words of length $n$ into $W_n;$ we denote corresponding mapping via
$\mathfrak{F}_{\mathfrak{A};n}(s).$ It is clear now that behaviour of the automaton can be described in terms
of the mappings $\mathfrak{F}_{\mathfrak{A};n}(s)$ for all $s \in S$ and all $n \in \mathbf{N} = \{1, 2, 3, . . .\}.$  If all states of
the automaton are reachable from the state $s_0,$ it suffices to study only the mappings
$\mathfrak{F}_{\mathfrak{A};n}(s_0)$ for all $n \in \mathbf N. $ 

Of course, here ``suffices'' has the epistemic meaning, i.e., 
by considering the mappings $\mathfrak{F}_{\mathfrak{A};n}(s_0),$ instead of the original automaton, we neglect by the  
dynamics of the ontic states; in our case the dynamics of conformational states. We also make the following methodological comment. 
The ideology of representation of an automaton by mappings $\mathfrak{F}_{\mathfrak{A};n}(s_0)$ is different from the ideology 
presented in this paper: transition from the ontic automaton to the epistemic one through factorization of the ontic state space. 
By moving to the $\mathfrak{F}_{\mathfrak{A};n}(s_0)$-functional representation we neglect  the state dynamics.    

The system of functions $\mathfrak{F}_{\mathfrak{A};n}(s_0)$ can be considered as the mapping in the space of infinite words in the alphabet 
$F_p.$ The latter can be identified with the ring of $p$-adic integers $\mathbf{Z}_p.$  Thus each automaton can be represented by a mapping 
 $u= \mathfrak{F}_{\mathfrak{A}}: \mathbf{Z}_p \to \mathbf{Z}_p, u(x)= \sum_{k=0}^n \delta_k(x) p^k,$ where $ x= \sum_{k=0}^n x_k p^k$  and $\delta_k, k=0,1,2...,$ are 
``coordinate functions'' valued in $F_p.$  

Mappings generated by automata have very special structure. This structure is encoded in the coordinate functions. 
Here $\delta_k$ depends only on the coordinates $x_0,..., x_k$ of the variable $x: \delta_k= \delta_k(x_0,..., x_k).$ In the protein model this means that 
protein cannot feel future states of the environment; its state at the instance of time $t_k=k$ depends only on environment's states at the previous instances 
of time $t=0,1,...,k.$  It can be easily shown that such  function $u= \mathfrak{F}_{\mathfrak{A}}:  \mathbf{Z}_p \to \mathbf{Z}_p$ satisfies 1-Lipschitz condition:
\begin{equation}
\label{L1} 
\rho_p(u(x), u(y))\leq \rho_p(x, y), x,y \in \mathbf{Z}_p,
\end{equation}
where $\rho_p$ is the $p$-adic distance.
This class of functions in relation to theory of automata was introduced by V. Anashin \cite{A1}, \cite{A2}, \cite{A3}, \cite{A4a}, \cite{A4} and it plays the important role in theory 
of $p$-adic dynamical systems, e.g.,  \cite{vander2, vander3}.
We call the space of  1-Lipschitz functions  the {\it Anashin functional space.} Thus in the epistemological model one can in principle consider just a mapping   $\mathfrak{F}_{\mathfrak{A}}$
belonging to the Anashin space (if we are not interested in the dynamics of epistemic states). The following natural question arises:

{\it Can any mapping belonging to Anashin's space be  generated by some automaton?} 

The answer is `yes', but in general the state space of the automaton  can be infinite \cite{A4a}.   As we have already discussed, by treating protein-automata as computational machines
we have to restrict modeling to finite state spaces. At the same time physics gives us arguments in favor of infinite and even continuous state spaces.

The problem of description of functions representing finite automata in terms of generated 
$p$-adic mappings were solved  by   Vuillemin \cite{V} (for $p=2$),  Smyshlyaeva \cite{SM} and Anashin \cite{A4} (for an arbitrary $p$). Their solutions, although beautiful mathematically (and based on the use of coordinate functions and 
 van der Put series, respectively), are too technical and too abstractly formulate to be  applicable  to 
our biological modeling. 

Now we discuss other special features of automata modeling  of proteins' behavior (in Anashin-like manner, i.e., by using the representation by $p$-adic functions).
We have already pointed out that the Moore machines seems to be more natural for such modeling than the general Mealy machines. However, in the $p$-adic functional 
approach the difference between these two types of finite states machines disappears: the classes of functions representing these machines coincide. For  any function representing 
a Mealy machine, it is possible to construct a Moore machine representable by this function (of course, the inverse statement is also true). This feature of Anashin's representation
is not so natural for our approach to proteins' modeling. However, the real situation is more complicated, since we are not sure that the space of protein's conformational 
states can be represented adequately as a finite state space. Thus it may be that we have to proceed with ``generalized Moore machines'' having infinite state spaces. 
We do not know whether functional classes representing such machines and general automata coincide or not.

The aforementioned result about the coincidence of $p$-adic functional classes for the Moore and Mealy machines shows that the functional approach, although very powerful 
from the mathematical viewpoint, shadows the internal structure of an automaton. We would like to mention another problem which is especially important for non-finite machines.
The state space of a biological or physical system is not only a set of point; typically it is endowed with some topological structure which plays the important role in understanding 
of functioning of this system. The simplest examples are state spaces $S=\{s_1,..., s_n,...\}$ with the discrete topology and $S=\mathbf R$ with its continuous topology. 
It seems that this difference in topological structures of state spaces disappear in the functional representation. However, may be a deeper analysis can clarify this connection 
between topology on $S$ and the special features of the corresponding $p$-adic mapping $\mathfrak{F}_{\mathfrak{A}}.$ 
For example, suppose that $S=\mathbf Z_p$ and that $f:  \mathbf Z_p \times \mathbf Z_p \to  \mathbf Z_p$ 
and $g:  \mathbf  Z_p \times \mathbf Z_p \to  \mathbf Z_p$ are continuous functions. May be the corresponding mapping      
$\mathfrak{F}_{\mathfrak{A}}:  \mathbf{Z}_p \to \mathbf{Z}_p$ has some special features? May be the difference can be monitored not at the level of a topological structure, but 
a differential structure on $S?$ For example,   $\mathfrak{F}_{\mathfrak{A}}:  \mathbf{Z}_p \to \mathbf{Z}_p$ is differentiable if both $f$ and $g$ are differentiable.

We continue the discussion on determination of the subspace of the Anashin functional space corresponding to ``protein automata.''
In general theory of automata one of the commonly used classes is the class of {\it transitive automata.}
 We recall that a family $U$ of mappings of a finite non-empty set $M$ into $M$ is called transitive whenever given a pair
$(a, b) \in  M \times M,$ there exists $u \in F$ such that $u(a) = b.$ An automaton is transitive if it can generate an arbitrary output from an arbitrary input. Anashin \cite{A4a} characterized
transitive automata in terms of ergodicity of corresponding mappings from  $\mathbf{Z}_p$ to  $\mathbf{Z}_p.$  
Since for our protein-modeling we are interested in finite-automata, we  make two remarks on interrelation of ergodicity and the possibility of the finite-automata realization  \cite{A4a}, \cite{A4}:
\begin{itemize}
\item  There exist ergodic mappings belonging to the Anashin class which are not representable by finite automata. For example, ergodic polynomials of degrees $> 1$ cannot be represented 
by finite automata. 
\item An ergodic polynomial of the degree one, i.e., an affine mapping of the form  $f(x)=ax+b$ can be represented by a finite automaton if and only if its coefficients 
$a, b$ are rational $p$-adic numbers, i.e., they belong to the intersection of $\mathbf{Z}_p$ and $\mathbf{Q},$ where $\mathbf{Q}$  is the field of rational numbers.\footnote{ The number 
$x$ belonging to $\mathbf{Z}_p$ is rational if and only if its representation by the sequence of digits, see (\ref{STSPI}), is periodic, i.e., some block of digits is repeated periodically. This characterization 
is similar to characterization of rational numbers in the field of real numbers $\mathbf{R}:$ these are numbers having periodic decimal representations.}   
\end{itemize}
We remark that consideration of  mappings determined by rational parameters matches with the ideology of $p$-adic theoretical physics, see, for example, \cite{PHYS1}--
\cite{PHYS2}, \cite{KHR_P2}. Here only rational numbers are treated as 
``physical numbers'', numbers which can get experimental verification.  

However, this very useful class of automata, transitive-automata,  does not match 
with our protein model. As was discussed at the very end of section \ref{AMI}, a protein is very specialized machine and its reacts very concretely to any input pattern of the environment producing very special functional
output. Thus proteins have to be modeled by non-transitive automata. 

The simplest non-transitive automata are given by locally constant mappings, i.e., continuous mappings taking only finitely many values. We remark that in the $p$-adic and 
more generally ultrametric case the characteristic function of a ball is continuous! A locally constant mapping can be represented as a linear combination of such characteristic functions, i.e., 
$\mathbf{Z}_p$ can be represented as the disjoint union of  (finitely many) balls such that this mapping takes the constant value at each ball. Locally constant functions play the important role 
in the $p$-adic (ultrametric) analysis. They form the dense subspace in the space of all continuous functions. 

In section \ref{UM} starting with the ultrametric structure of the ontic state space we proposed representation of the epistemic state space as the space of balls. 
Therefore it is natural to assume that the output  function (observational function) $g$ is locally constant on the ontic state space $S.$ Therefore it is natural as well to suppose that the 
realization of the protein-automaton by a mapping  $\mathfrak{F}_{\mathfrak{A}}: \mathbf{Z}_p \to \mathbf{Z}_p$ is also a locally constant function. Now the following question 
arises: {\it Which locally constant functions can be realized by finite automata?} The answer is known \cite{A4a}: only functions taking the $p$-adic rational values. This again matches with 
the ideology of the $p$-adic theoretical physics.  

Under the assumption that behavior of proteins can be mathematically modeled with the aid of the Mealy machines (finite state automata),   we can guess 
that the class of protein-automata is a subclass of the class of automata 
for which $\mathfrak{F}_{\mathfrak{A}}$ is a locally constant $p$-adic rational-valued function. However, to describe this subclass precisely we have to analyze 
this problem taking into account more delicate biological and molecular features of representation of proteins by automata.   
In general  characterization of the class of automata matching proteins' behavior is the interesting  mathematical problem. 
And we hope that publication of this concept-paper would attract interest of mathematicians and molecular biologists to study of this problem.

\end{document}